\documentclass[useAMS,usenatbib,usegraphicx]{mn2e}

\usepackage{aas_macros}

\usepackage{color}

\def\gx{GX~339$-$4}

\def\1e{1E~1740.7$-$2942}

\def\xteonze{XTE~J1118$+$480}

\def\vq{V404~Cyg}

\def\a0{A~0620$-$00}

\def\1h{H~1743$-$322}

\def\4u{4U~1755$-$33}

\def\ledd{\hbox{L$_{\rm Edd}$}}

\def\ergs{\hbox{erg s$^{-1}$}}

\title[\vq\ in quiescence]{Revisiting the radio/X-ray flux correlation in the black hole \vq : from outburst to quiescence}
\author[S. Corbel et al.]{S. Corbel$^{1}$\thanks{E-mail:corbel@discovery.saclay.cea.fr} , 
E. Koerding$^{2}$ and P. Kaaret$^{1,3}$.\\
$^{1}$Universit\'e Paris 7 Denis Diderot and Service d'Astrophysique, UMR AIM,  CEA Saclay, F-91191 Gif sur Yvette, France.\\
$^{2}$School of Physics and Astronomy, University of Southampton, Highfield, Southampton, Hampshire, SO17 1BJ, UK.\\
$^{3}$Department of Physics and Astronomy, University of Iowa, Iowa City, IA 52242 USA.}

\begin{document}


\pagerange{\pageref{firstpage}--\pageref{lastpage}} \pubyear{2008}

\maketitle

\label{firstpage}

\begin{abstract}

We report results of {\em Chandra} X-ray and VLA radio observations of
the Galactic accreting black hole \vq\ (GS~2023$+$338) in its quiescent
state. \vq\ is detected at its faintest level of radio and X-ray
emission with a 0.5--10 keV unabsorbed luminosity of 8.3 $\times$
10$^{32}$ (d/3.5 kpc)$^2$ \ergs .  The X-ray spectrum fit with an
absorbed power-law model yields a photon index of 2.17 $\pm$ 0.13. 
Contrary to previous findings, this clearly indicates that \vq\
undergoes -- like most black holes in quiescence -- a softening of its
X-ray spectrum at very low luminosity compared to the standard hard
state.  The quiescent radio emission is consistent with the presence of
self-absorbed compact jets.  We have also reanalyzed archival data from
the decay of the 1989 outburst of \vq\ in order to quantify more
precisely the correlation between radio and X-ray emission in the hard
state of \vq.  We show that this correlation extends over five decades
in X-ray flux and holds down to the quiescent state of \vq .  The index
of this correlation ($\sim$ 0.5) may suggest that synchrotron
self-Compton emission is the dominant physical process at high energy in
\vq .  However, this index is also  consistent with scale invariant jet models 
coupled to an inefficiently radiating accretion disc. We discuss the properties 
of the quiescent state of black holes and highlight the fact that some of their 
properties are different from the standard hard state.

\end{abstract}

\begin{keywords}

accretion, accretion disks -- binaries: general -- ISM: jets and
outflows -- radio continuum: stars -- X-rays: stars -- stars: individual
(V404 Cyg).

\end{keywords}

\section{Introduction}

Accreting black holes in X-ray binaries are known to undergo transitions
between various ``X-ray'' spectral states (see \citet{McClintock06}  for
a review), mainly (but not only, see \citealt{Homan01}) due to variation
of the accretion rate within the accretion disk.  The quiescent state is the 
lowest luminosity state and is a factor of $\sim$10$^6$  or more fainter than 
the brightest outburst state. In addition, a hard state is usually observed in the
initial and final phases of an outburst with typical luminosity in the
range 10$^{-3}$--10$^{-1}$ of the Eddington luminosity.

The quiescent and hard states share similar properties  (e.g.
\citealt{Tomsick04}).  Indeed, the quiescent state is often viewed as a
lower luminosity version of the hard state. The compact jet observed in
the hard state \citep{Corbel00} appears also in quiescence as inferred
from the characteristics of the radio spectrum \citep{Gallo06}. The
strong correlation between radio and X-ray emissions in the hard state
\citep{Corbel03} seems to be maintained down to quiescence
\citep{Corbel03,Gallo03,Gallo06}.

However, current X-ray satellites (especially  {\em Chandra} and {\em
XMM-Newton}) have revealed new details of the spectrum of quiescent
black holes. It appears that a fraction of them display a softer X-ray
spectrum compare to the standard hard state \citep{Corbel06}. In
addition, deviations to the standard radio/X-ray flux correlation have
been observed in the black hole \gx\ at very low luminosity (Corbel et
al. in prep.). These peculiarities might imply that the quiescent state
has to be considered as distinct from the standard hard state.

The universal radio/X-ray flux correlation presented by  \citet{Gallo03}
is dominated by two sources (\gx\ from \citet{Corbel03} and \vq ) plus
additional points from other sources.  \a0\ in quiescence is,
remarkably, consistent with an extrapolation of the \vq\ and \gx\
correlations down to quiescence \citep{Gallo06}, but the exact track of
\a0\ in outburst is unknown as no radio observations of a hard state
were conducted in that time.  The correlation observed in the hard state
of accreting galatic black holes has been extended to active galactic
nuclei by  including  an additional correction for taking into account
the mass of the black hole \citep{Merloni03,Falcke04,Kording06}.  This
fundamental plane of black hole activity relies strongly on the
correlation observed in  \vq\ \citep{Gallo03} and \gx\
\citep{Corbel03}.  Thus, it is important to assess the reliability of
the correlation for Galactic systems.

\vq\ has the longest orbital period of any black hole system detected in
quiescence to date. \citet{Corbel06} reported that long orbital period
systems have quiescent spectra consistent with the hard state, contrary
to short orbital period systems that have softer spectra.  However, the
long orbital period group was statically dominated by the spectrum of
\vq\ and recent {\em XMM-Newton} observations of \vq\  show a soft
quiescent spectrum \citep{Bradley07}.  Therefore, we used a new {\em
Chandra} observation to reconsider the X-ray spectrum of \vq\ and the
properties of black holes in quiescence.

In this paper, we describe the results of radio and X-radio observations
of \vq\ in quiescence  and re-examine archival observations of \vq\
during the decay of its 1989 outburst (\S 2). The observations in
quiescence provide a detailed measurement of  the X-ray spectrum of
\vq\  as well as a radio detection at its  faintest level of emission 
(\S 3).  These first simultaneous radio and X-ray observations of  \vq\
in quiescence led us to revisit the radio/X-ray flux correlation in
\vq\  (\S 4). For that purpose, we reconsidered all X-ray observations
of \vq\  during its 1989 outburst, allowing us to study this correlation
in much finer detail. Our conclusions are summarized in section \S 5.

\section{Observations and data analysis}

\subsection{Chandra observation of \vq\ in quiescence}

\vq\ has been observed by  {\em Chandra} on two occasions while it was
in or close to quiescence. A first $\sim$ 10~ks observation performed in
2000 has been published  by \citet{Kong02}, while the second one
conducted in 2003 has  been analyzed by \citet{Hynes04}.  However,
\citet{Hynes04} only reported on the short term X-ray variability and
contemporaneous optical behavior.  Because no spectral analysis of the
2003 observations has been published, we decided to re-analyze these
data in light of the possible difference in X-ray photon indices between
short and long orbital period black hole systems suggested in
\citet{Corbel06}.

In 2003, \vq\ was observed by {\em Chandra} on July 28 and 29 for $\sim$
60~ks with the Advanced CCD Imaging Spectrometer spectroscopic array
(ACIS-S; \citealt{Bautz98}).  \vq\ was placed on the back-illuminated
ACIS Chip S3 in 1/8 sub-array mode to reduce pileup. We constructed
light-curves from all valid events on the S3 chip to search for times of
high background. We found no background flares and therefore use the
whole {\em Chandra} observation for the spectral analysis giving an
exposure of 55.6~ks. A standard data analysis was
performed using the {\em Chandra} Interactive Analysis of Observations
(CIAO) software package -- version 3.4.1.1 -- with the most  up to date
calibration database -- version 3.4. 

A bright X-ray source is detected at the location of \vq . We extracted
the energy spectrum of \vq\ using a circular source extraction region of
3\arcsec , providing a total of 1943 photons. We rebinned the  spectrum
to have at least 30 counts in each energy bin. As already noted by
\citet{Hynes04}, the 2003 observation was performed when \vq\ had on
average a factor 5 lower count rate compared to the 2000  {\em
Chandra}  observation reported by  \citet{Kong02}.  A background spectrum 
was extracted from an annulus with an inner radius of 4\arcsec\ and an outer
radius of 18\arcsec .  However, given the brightness of  \vq , this
background is negligible.

\subsection{VLA observation of \vq\ in quiescence}

We have reduced archival VLA data (PI: Hynes, AH823) of \vq\  in
quiescence. The VLA observation in A-configuration was conducted on the
29th July between UT 0:20 and 14:15, simultaneously with the Chandra
observation.  The total time on source was $\sim$5~hours at 8.4~GHz  and
4.4~hours at 4.8~GHz. The target was observed using phase-referencing to
the secondary calibrator 20252+33430.  The absolute flux calibration was
established by observing the quasar 3C 286.  We assume that the absolute
flux calibration introduces a systematic uncertainty of $\sim$ 5 \%. 
The data reduction was performed using standard AIPS procedures.

To our knowledge, these VLA observations are the only radio observations
that have been conducted strictly simultaneously with X-ray observations
while \vq\ was in a quiescent state. The quiescence observations
reported in \citet{Gallo03} were not simultaneous.  As \vq\ is known to
vary significantly in its quiescent state \citep{Hynes04,Bradley07},
strict simultaneity is important to precisely contrain the X-ray versus
radio flux correlation.

\subsection{The 1989 outburst of \vq\ : archival data }

In late May 1989, {\em Ginga} detected new activity from GS~2023$+$338
consistent with the location of the variable star \vq . Analysis of the
X-ray outburst initially indicated that the source stayed in the hard 
state for the whole duration of its active period \citep{Oosterbroek97}.
According to \citet{Zycki99}, a very short excursion to a thermal state
was observed early in the outburst. However, since a bright radio flare
was detected by \citet{Han92} very early in the outburst and such bright
ejection events are usually associated with state transitions 
\citep[e.g.][]{Corbel04b, Fender04b}, it seems very likely that \vq\ was
in a different X-ray state in the early phase of the outburst and that
the ejection event was associated with a state transition.

We are primarily interested in precisely locating \vq\ in quiescence on
the radio/X-ray correlation using the simultaneous X-ray and radio
fluxes reported below.  We therefore decided to see if we could
improve the original work of \citet{Gallo03} on the correlation in \vq.
For this purpose, we used the original X-ray data from the 1989 outburst
as measured by {\em Ginga} (LAC and ASM), as well as by {\em MIR/Kvant}
(HEXE and TTM).  These data were available in form of flux light-curves
(in erg s$^{-1}$  cm$^{-2}$) in different energy bands (see
\citealt{Chen97} for more details).  We assume an error of 10\%  for
the  {\em MIR/Kvant} fluxes as no errors are quoted for these
light-curves. We considered only the portion of the light-curves during
the decay when \vq\ is clearly in a standard hard state with a typical
inverted radio spectrum as reported by \citet{Han92}, specifically, we
use only data after 1989 June 8.

\begin{figure}
\includegraphics[width=84mm]{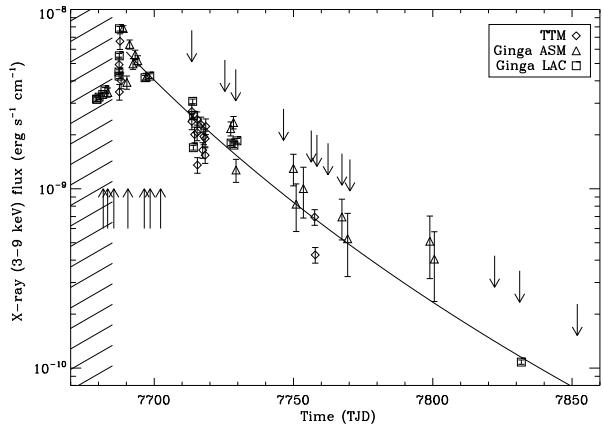}
\caption{X-ray lightcurve in the 3--9 keV band constructed from archival
data of {\em Ginga} (LAC and ASM) and {\em MIR/Kwant} (TTM).  The line
is a power-law fit to the decay of the outburst and has been used to
estimate the X-ray flux during the radio observations.  Arrows indicate
the dates of the radio (VLA) observations of \citet{Han92}. The radio
data in the dashed area have not been used in our study.  }
\label{fig_xlc}
\end{figure}

We first constructed light-curves (Figure \ref{fig_xlc}) in the standard
X-ray band (3--9~keV) used for these correlation studies by converting
the original flux data in the 3--9~keV band using WebPimms from
HEASARC\footnote{http://heasarc.nasa.gov/Tools/w3pimms.html} assuming a
shape for the X-ray spectrum typical of the hard state, specifically a
power-law with photon index of 1.6, \citep{McClintock06} and a
hydrogen column density of $8 \times 10^{21} \rm \, cm^{-2}$.  We note
that extra absorption was observed during outburst
\citep{Oosterbroek97,Zycki99,Zycki99b}, but its effect is not
significant in  the 3-9 keV band used in this study.  At this stage, we
do not use X-ray fluxes in Crab units as the Crab spectrum is
significantly different from a typical hard state spectrum \citet{Gallo03}.

By interpolating the X-ray decay light-curve (Figure \ref{fig_xlc}), we
were able to obtain reasonable estimates for the 3--9~keV flux at the
times of the radio observations listed in Table 1 of \citet{Han92} (ten
of these observations were quasi-simultaneous with an X-ray
observations).  In addition, when \vq\ was very faint (1989  November 1
and 1990 August 16), we reduced directly the {\em Ginga} LAC spectrum
provided by HEASARC in order to obtain a very good estimate of the
X-ray flux.  In total, we have 20 pairs of measurements/estimates of the
radio and X-ray flux of \vq\  during the decay of its 1989 outburst
(compared to 10 measurements in \citealt{Gallo03}).  We do not use the
quiescence point of \citet{Gallo03} because the X-ray and radio
observations were not conducted at the same time, but we use, instead,
our quiescence measurements as describe in this paper.  In addition, our
procedure led to another important quasi-simultaneous measurement at an
X-ray flux almost two decades brighter than quiescence (see the point
labeled Ginga/VLA in  Figure 4).  These measurements should be more
precise than those in the broader study of \citet{Gallo03}.

\section{Results}

\subsection{The X-ray spectrum of \vq\ in quiescence}

Even though \vq\ is the brightest quiescent black hole
\citep[e.g.][]{Tomsick03}, the {\em Chandra} spectrum is adequately 
well fitted with a simple phenomenological model consisting of a
power-law modified by interstellar absorption.  This model is adequate
for our purpose, as we aim primarily to compare the power-law photon
index of \vq\ with the values catalogued in \citet{Corbel06} for a
sample of quiescent black holes (see also \citealt{Kong02}).   More
detailed spectral modeling of \vq\ with {\em XMM-Newton}  can be found
in \citet{Bradley07}.

\begin{figure} \includegraphics[width=54mm,angle=270]{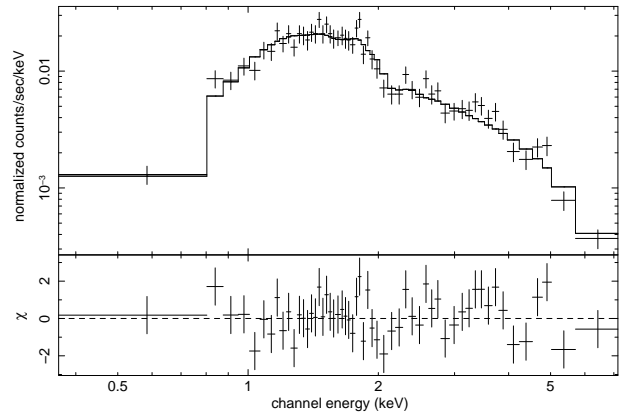}
\caption{ {\em Chandra} energy spectrum of \vq\ in 2003 from 0.3 to
8~keV fitted with a power-law modified by interstellar absorption. The
residuals in the form of the contribution to the value of $\chi^2$ for
each energy bin are plotted in the bottom panel. } \label{fig_spec}
\end{figure}

Fitting the {\em Chandra} spectrum (Figure \ref{fig_spec}) with the
absorbed power-law resulted in a photon index $\Gamma$  of 2.17 $\pm$
0.13 and a hydrogen column density  $N_\mathrm{H}$ of (8.1 $\pm$ 0.1)
$\times$ 10$^{21}$~cm$^{-2}$ and gave an adequate fit with $\chi^2$ =
62.3 for 53 degrees of freedom.  The quoted uncertainties are at the
90\%  confidence level for one parameter.  The hydrogen column density 
deduced from this {\em Chandra} observations is slightly greater than
the one deduced from optical measurement, but this is not unusual as
already mentioned by \citet{Kong02}.   These results are fully
consistent with the spectral parameters measured for the {\em
XMM-Newton} observation conducted in November 2005 ($\Gamma$ = 2.09
$\pm$ 0.08) as reported by \citet{Bradley07}. The unabsorbed 3-9 keV 
flux during the 2003 {\em Chandra} observation is 
(1.79$^{+0.13}_{-0.06}$) $\times$ 10$^{-13}$ erg s$^{-1}$ cm$^{-2}$. 
This corresponds to a 0.5--10 keV unabsorbed luminosity of 8.3 $\times$
10$^{32}$ (d/3.5 kpc)$^2$ \ergs .  This is a factor 5 fainter than in
the previous  {\em Chandra} observation conducted in 2000 
\citep{Kong02}, consistent with the flux in 2005 reported by
\citet{Bradley07}, and slightly fainter than in the {\em BeppoSAX}
observations \citep{Campana01}.

\begin{figure} \includegraphics[width=84mm]{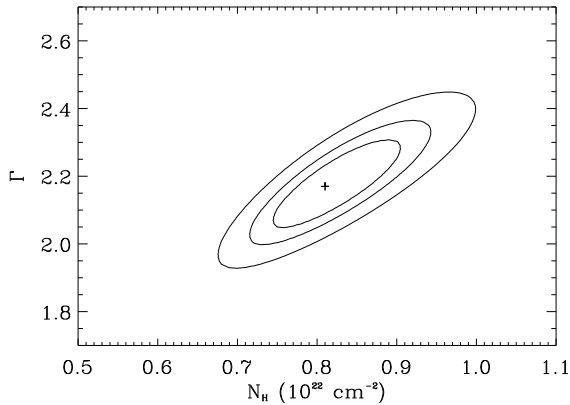} \caption{Error
contours for the hydrogen column density (N$_\mathrm{H}$) and the
power-law index ($\Gamma$) derived from the {\em Chandra} spectrum of
\vq. The cross marks the location of the best-fit value, and 68\% ($
\Delta \chi^2$ = 2.30), 90\% ($ \Delta \chi^2$ = 4.61) and 99\% ($
\Delta \chi^2$ = 9.21) confidence contours are shown.} \label{fig_cont}
\end{figure}

Figure \ref{fig_cont} illustrates the 68\%, 90\%, and 99\% error
contours allowing two parameters ($\Gamma$ and N$_\mathrm{H}$) to vary,
and indicates that these parameters are well constrained by our
observations. The {\em Chandra} and the  {\em XMM-Newton} observations
therefore confirm that the X-ray spectrum of \vq\ is softer in
quiescence than in outburst, contrary  to what was originally reported
by \citet{Kong02} from the 2000 {\em Chandra} observations
 (they obtained a photon index of 1.55 $\pm$ 0.07  with an hydrogen column 
 density frozen to the optical measurement, and 1.81 $\pm$ 0.14 with 
 N$_\mathrm{H}$ free to vary). The new  {\em Chandra} and {\em XMM-Newton}  data 
 are both consistent in term of N$_\mathrm{H}$ estimations and favour a value 
 higher that the one deduced from optical observations.
In addition, we found that the 2000 {\em Chandra} spectrum of \citet{Kong02} was
moderately  affected by pile-up. We refitted their spectrum using the pile-up
model in {\em Xspec} and found $\Gamma$ = 2.06$^{+0.26}_{-0.29}$ with 
N$_\mathrm{H}$ = 7.5$^{+1.5}_{-0.8}$ $\times$ 10$^{21}$~cm$^{-2}$, values 
that are now fully consistent with the new  {\em Chandra} and  {\em XMM-Newton} spectra.

\subsection{Radio emission from \vq\ in quiescence}

A radio source is detected at the location of \vq\ with flux densities
of 164 $\pm$ 38 $\mu$Jy and 193 $\pm$ 22 $\mu$Jy at 4.8~GHz and 8.4~GHz
respectively, giving a spectral index of $\alpha$ = +0.29 $\pm$ 0.46 
defined as $S_\nu$ $\propto$ $\nu^\alpha$ where $S_\nu$ is the flux
density at frequency $\nu$.  This is the faintest level of radio 
emission ever reported from this source.  In light of the significant
X-ray variability during the {\em Chandra} observations \citep{Hynes04},
we searched for radio variability by dividing the radio dataset in two
parts, but we did not find any significant radio variabilty.

Despite a fainter level of emission, the radio spectrum of \vq\
(admittedly not very well constrained) is fully consistent with previous
observations conducted with Westerbook \citep{Gallo05b}.  A flat
spectrum and stable radio emission are fully consistent with the
presence of self-absorbed compact jets \citep{Hjellming88,Blandford79} 
in the quiescent state of \vq\ (see also the discussion in
\citealt{Gallo05b}). In addition, the Spitzer data of \citet{Gallo07}  would still
be consistent with an extrapolation of our new radio spectrum up to infrared 
and therefore our results do not alter their conclusions.

\section{Discussion}

\subsection{Softening of black holes X-ray spectra in quiescence}

By analyzing the X-ray spectra of black holes in quiescence,
\citet{Corbel06} found that a significant fraction of those black holes
had an X-ray spectrum that was significantly softer in the quiescent
state with respect to the brighter standard hard state. They also
noticed that the three black holes with the highest orbital periods had
spectra consistent with the hard state, whereas the short orbital period
($<$ 60 hours) systems were all consistent with a soft spectrum.
However, the difference between the two groups was statically dominated
by the 2000 {\em  Chandra} spectrum of \vq .

As described above, the 2003 {\em Chandra} and 2005 {\em XMM-Newton} 
spectra of \vq\ clearly indicate that \vq\ also undergoes a softening in
quiescence. These results differ from the conclusion of
\citet{Corbel06}.  That conclusion was based on the assumption that the
2000 {\em Chandra} spectrum of \vq\ \citep{Kong02} represented the
quiescent state of \vq\ (which we have now shown in section 3.1 is
also consistent with the new measurements if pile-up is taken into account). 
In light of our new results, we find that
there is no statistically significant difference in the quiescent X-ray 
spectra of black holes with long or short orbital period.  On the
contrary, the new spectrum of \vq\ generalizes the finding of 
\citet{Corbel06} that {\it all black holes in quiescence have a softer
X-ray spectrum than the standard hard state.}

As discussed in \citet{Corbel06}, there are several different possible
physical reasons for this softening.  The softening could be related to
an advection dominated accretion flow (ADAF) \citep{McClintock03} or a
jet located above an ADAF \citep{Yuan05}.  However as outlined in recent
studies (e.g. \citealt{Gallo06}), ADAF solutions have several 
difficulties in self-consistently modeling the powerful outflows that
may be present in quiescence.

Interestingly, now that the softening of \vq\ is clearly brought to
light, inspecting the original {\em Ginga} LAC data revealed that such
softening was already on beginning a few months after the peak of the
1989 outburst.  Indeed, after a peak around 1989 May 30, the  {\em
Ginga}  spectrum in November 1989 was much softer than in the other 
observations conducted during the decay  (as indicated by Table 2 in 
\citealt{Zycki99b}).  The transition to the soft quiescent state
therefore occurred around a 0.5--10 keV unabsorbed flux of 2.5 $\times$
10$^{-10}$ erg s$^{-1}$ cm$^{-2}$ equivalent  to a luminosity of $\sim$
3 $\times$ 10$^{-4}$ $\ledd$ for a distance of 3.5 kpc (according to
\citealt{Zycki99b}).  In addition, it is worth noting that other black
holes have also shown indication of softening at these intermediate
luminosities (see  section 4.3 in \citealt{Corbel06} and  references
therein).

Black hole systems at luminosity below 10$^{-4}$ to 10$^{-5}$ $\ledd$
may, therefore, enter a quiescent state with different properties (at
least a softer X-ray spectrum) than the hard state.  This is contrary
the standard view that the quiescent state is an extension of the hard
state, which has arisen due to the continuous evolution of some of the
X-ray properties (e.g.\ timing) at low luminosity  (e.g.\
\citealt{Tomsick04}). The quiescent state may be related to new
accretion disk properties or other characteristics of the outflow, e.g.\
inefficient particle acceleration.  Further observations of black holes
during the decay of their outbursts would (as in \citet{Kalemci06})  be
important to constrain the properties of the quiescent state.

\subsection{Revisiting the radio/X-ray flux correlation in \vq }

\vq\ was the second galactic black hole for which a strong correlation
between radio and X-ray emission was observed \citep{Gallo03}.  As
outlined in section 2.3, we have reanalyzed some of the data from its
1989 outburst, allowing us to obtain a more precise estimate of the
X-ray flux during the decay of the outburst, as well as almost double
the number of pairs of near simultaneous radio/X-ray fluxes.  With the
2003 Chandra/VLA observations and the 2000 Ginga/VLA observations, we
are able to provide two precise and simultaneous fluxes while \vq\ was
in quiescence.  This allow us to revisit the correlation found by
\citet{Gallo03} in much finer detail than before.

\begin{figure} \includegraphics[width=84mm, angle=0]{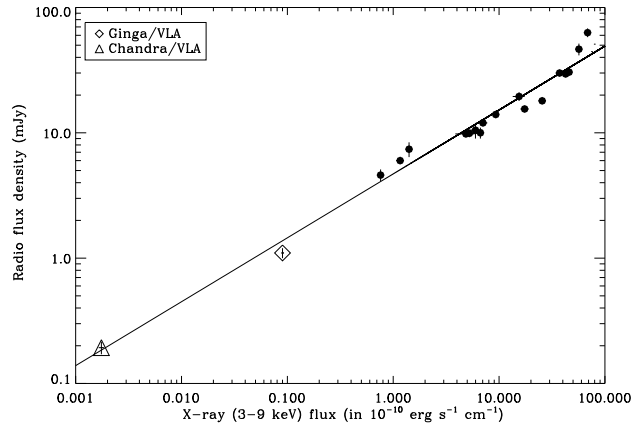}
\caption{Radio flux density F$_\mathrm{Rad}$ at 8.4 GHz versus the
unabsorbed 3-9 KeV flux $F_\mathrm{X}$ for \vq\ for the decay of the
1989 outburst and the recent {\em Chandra/VLA} observations in
quiescence.  The straight line is a fit to these data-points with a
function of the form $F_\mathrm{Rad}$ $\propto$ F$_\mathrm{X}^b$, with
$b$ = 0.51 $\pm$ 0.06.  The two points with the lowest fluxes were not
included in the fit, but they  agree perfectly with the correlation
observed in the decay of 1989 outburst.} \label{fig_correl} \end{figure}

For that purpose, we only consider the radio observations after TJD 7685
(1989 June 8) as a secondary X-ray and optical re-flare was observed 
during the decay (see light-curves in \citealt{Brocksopp04}). This
ensures that we are considering observations that are in the hard state,
as confirmed by the flat or inverted radio spectrum after this date that
is a characteristic of the self-absorbed compact jets
\citep{Blandford79}.  In figure \ref{fig_correl}, we plot the radio flux
density $F_\mathrm{Rad}$ at 8.4~GHz versus the unabsorbed 3--9 keV flux
$F_\mathrm{X}$, indicating (as originally observed for \vq\ by
\citealt{Gallo03}) that a strong correlation between these two
frequency  domains is present.

We fit the correlation with a power law of the form $F_\mathrm{rad} =
k~F^b_X$ in log-space, taking the uncertainties in both the measured
radio and X-ray flux into account.  A fit to all the data with the
measured uncertainties yields $k = 4.6 \pm  0.3$ and $ b = 0.51 \pm
0.02$, with a merit function of 102 for 17 degrees of freedom. Thus,
there are some deviations from the power law fit exceeding the
uncertainties.  We can parametrise these deviations by adding some
isotropic excess scatter to the data.  To obtain a good fit we have to
introduce 0.068 dex scatter. With the excess scatter, we find for the
fit parameters $k = 4.6 \pm  0.4$ and $b = 0.51 \pm 0.03$ (in log
space,  we obtain log $F_\mathrm{rad}$ = (0.66 $\pm$ 0.04) log
$F^b_X$).  If we fit the data without the last two points near quiescence
we find the same fit values albeit with larger uncertainties: $k = 4.7
\pm  0.6$ and $b = 0.51 \pm 0.06$.  The quiescent measurements are
consistent within the uncertainties with the extrapolation of this fit
to the luminous data.  A fit using the radio fluxes at 4.8 GHz  from
\citet{Han92} instead of 8.4 GHz does not change the overall index, as
one obtain $k =4.0 \pm  0.4$ and $b = 0.53 \pm 0.04 $.

To test if significant spectral evolution of the source is responsible
for the low correlation index (0.51 compared to 0.7), we also fit the
correlation using 1.2 to 37.2 keV X-ray fluxes. Here, the scatter is
slightly larger (0.14 dex excess scatter) but we find similar fit values
as before $k = 1.94^{+0.93}_{-0.63}$ and $ b = 0.52 \pm 0.10$.  The
larger uncertainties mainly arise due to the lower sample size for which
1.2 to 37.2~keV fluxes are available (only 8 measurements). We therefore
do not see an evolution of the correlation index in the 1.2 to 37.2~keV
energy range.

The correlation observed during the decay of the 1989 outburst therefore
holds down to deep in the quiescent state.  Contrary to the behavior
observed from \gx\ (Corbel et al.\ in prep), there is no deviation to
the correlation for \vq\ in quiescence, at least when comparing to the
track measured from the 1989 outburst (several tracks are observed in
\gx ; Corbel et al.\ in prep.). This constitutes the best example
(almost five decades in X-ray flux) -- to date -- for a non-linear
correlation between radio and X-ray fluxes in the hard state of a black
hole candidate.  We note that even if the dispersion of the data-points
along the fit in Figure \ref{fig_correl} is much smaller than in
\citet{Gallo03}, there is still some fluctuation (quantified with the
isotropic excess) along the fitted line (in log space).  This residual
deviation may possibly be related to our procedure of estimating the
X-ray fluxes of \vq\ (c.f. section 2.3) or possibly to variation in the
absorbing column density as observed by {\em Ginga} 
\citep{Oosterbroek97,Zycki99,Zycki99b}.  The latter is more likely as
some fluctuations were also observed at radio frequencies 
\citep{Han92}.

Our measured power law index of $b = 0.5$ is significantly lower than
the usual value of $b \sim 0.7$   \citep{Corbel03,Gallo03} (but it is 
consistent with the value of 0.7 $\pm$ 0.2 measured by \citet{Gallo03}
using only data from \vq ). Our larger sample allow us to obtain 
a more precise estimation of this  correlation index,  and give us an 
opportunity to look in finer detail  into the physical meaning of the
correlation.  For example, this index is still consistent with scale invariant 
jet models coupled to an inefficiently radiating accretion disc \citep{Merloni03}.

Additionally, \citet{Yuan05b} adapted an ADAF+jet model, originally developed for the
hard state of \xteonze , to black holes in quiescence.  Assuming no
change in jets physics from the hard state to quiescence,
\citet{Yuan05b} would have expected a change in the correlation index
from b = 0.7 to  $\sim$ 1.23 below an X-ray luminosity of 10$^{-6}$ 
$\ledd$ , which is slightly above the quiescent luminosity of $\vq$  .
We do not observe this behaviour, which either implies that the ADAF+jet
model of \citet{Yuan05b} is not applicable to the quiescent state or
that jet  physics  change significantly in quiescence (a possibility
discussed in section 4.1).

In case that the X-ray emission originates from the jet, the low
correlation index may indicate that the X-ray emission is created by
synchrotron self-Compton emission (SSC) as also suggested by 
\citet{Markoff05}.  If the X-ray emission is SSC, then it depends more
strongly on the jet power ($\sim Q_{jet}^{\frac{11}{4}}$) than if it is
synchrotron emission ($\sim Q_{jet}^{\frac{7}{4}}$) (e.g.,
\citealt{Falcke95}).  Thus, the correlation between radio luminosity
(which depends on the jet power as ($\sim Q_{jet}^{\frac{17}{12}}$)) and
X-ray luminosity from SSC is flatter and gives $b \approx \frac{17}{12}/
\frac{11}{4} \approx 0.51$, a value fully consistent with our derived
index.  If the SSC interpretation is correct, then this is different
from the 1997-99 tracks of \gx\ which was consistent with $b$ = 0.7 and
a pure synchrotron origin \citep{Corbel03, Markoff03}.  However, a lower
correlation index was also noticed for \gx\ at intermediate luminosity
in the recent outbursts (Corbel et al.\ in prep).  This could 
potentially indicate that we are witnessing an interplay between various
emission processes (synchrotron, SSC, external Comptonisation) at high
energy, for reasons that still need to be understood.  In any case, 
a broadband fit to the spectral energy distribution  of A~0620$-$00, using the jet model of 
\citet{Markoff05},  is consistent  with an SSC origin for the hard X-ray 
emission in this quiescent
black hole.  It would be worthwhile to check for the presence of 
an SSC component in the broadband fit of \vq\, as implied by  our correlation study.

\section{Conclusions}

We publish the results of a {\em Chandra} and VLA observation of \vq\ in
quiescence and we reevaluate the correlation between radio and X-ray
emission.  The main conclusions of our work can be summarized as
follows:

\begin{itemize}
\item The {\em Chandra} and VLA observations have allowed a detection of
\vq\ at its faintest level of emission yet observed.  The
characteristics of the radio emission remain consistent with the
presence of self-absorbed compact jets in the quiescent state.

\item The {\em Chandra} observation of \vq\ confirms the softening of
its X-ray spectrum in quiescence.  This implies that all black holes in
quiescence have a softer X-ray spectrum in quiescence than in the 
standard hard state.

\item We revisit and improve the correlation between radio and X-ray
emission in \vq\ and found an index of the correlation that may suggest
that the X-ray emission in the hard state of \vq\ is due to synchrotron
self-Compton emission.  Compared to other sources, this would imply that
the X-ray emission in black holes could be the results of interplay of
various emission mechanisms (synchrotron, SSC or external 
Comptonisation) with different mechanisms favoured under conditions that
still need to be understood.
\end{itemize}

\section*{Acknowledgments}

EK acknowledges funding of a Marie Curie IEF under contract number
MEIF-CT-2006-024668.  The VLA is operated by Associated Universities,
Inc., under contract with the National Science Foundation.  This
research has made use of data obtained from the High Energy Astrophysics
Science Archive Research Center (HEASARC), provided by NASA's Goddard
Space Flight Center. SC would like to thank Piotr Zycki and Chris Done 
for providing the Ginga/LAC data of \vq\ used in
\citet{Zycki99,Zycki99b} and Shunji Kitamoto for providing the Ginga/ASM
data of \vq. SC also thanks Elena Gallo for providing the data used in
the original study of the  correlation in \vq , Sera Markoff for useful
discussions and J\'erome Rodriguez for discussion of the spectrum of the
Crab.

\bibliographystyle{mn2e}

\bibliography{Reference2008Feb13.bib}

\label{lastpage}

\end{document}